\documentclass[varenna]{cimento}
\usepackage{graphicx}

      [1999/12/01 v1.4c Il Nuovo Cimento]

\begin{document}

\title{Importance of Magnetism in Phase Stability, Equations of State, and
Elasticity}
\author{R.~E.~Cohen \atque S.~Gramsch}
\institute{Geophysical Laboratory, Carnegie Institution of Washington, Washington, D.C. 20015}
\author{G.~Steinle-Neumann \atque L.~Stixrude}
\institute{Department of Geological Sciences, University of Michigan, Ann Arbor, MI 48109}

\maketitle

\begin{abstract}
The effects of magnetism on high pressure properties of transition metals and transition metal 
compounds can be quite important. In the case of Fe, magnetism is responsible for stability of the 
body-centered cubic (bcc) phase at ambient conditions, and the large thermal expansivity in 
face-centered cubic (fcc) iron, and also has large effects on the equation of state and elasticity of 
hexagonal close-packed (hcp) iron. In transition metal oxides, local magnetic moments are responsible 
for their insulating behavior. LDA+U results are presented for CoO and FeO, and predictions are made 
for high pressure metallization. The inclusion of a local Coulomb repulsion, $U$, greatly inhibits 
the high-spin low-spin transitions found with conventional exchange-correlation functionals (i.e. 
generalized gradient corrections, GGA). We discuss theory and computations for the effects of 
magnetism on high pressure cohesive properties.
\end{abstract}

\section{Magnetism}

One might think that one needs to be concerned with magnetism only if one is
interested in magnetic properties of materials. This is not the case! Rather
the stable crystal structure can depend on properly taking into account
magnetism, which can strongly affect phase stability, lattice distortions, elasticity, equations of 
state, and vibrational frequencies. Furthermore, magnetism can be important in materials
containing transition metal ions even when the magnetic moments are not
ordered. The distinction between ``non-magnetic'' materials and disordered, 
paramagnetic, materials is crucial.

Magnetism in crystals comes about because electrons have a quantity called
spin, which is a vector quantity that behaves like quantum angular momentum. Electrons also have 
orbital angular
momentum, which can lead to orbital magnetic moments. In heavy transition
metal ions with partially occupied f-electron shells, such as rare earths or
lanthanides, these orbital moments can be very important.

Here we will concentrate on d-electron systems, but there is much complex and fascinating physics in 
the f-electron systems. In the f-electron systems, the f-electrons tend to form
very localized, strong magnetic moments on each atom. The localized
f-electrons then interact with delocalized band-like states, and this
interaction can lead to interesting phase transitions with pressure, for
example in Ce \cite{825}. Electronic d-states tend to be more delocalized than
f-states, though localization is important, as we will discuss below.

The fact that electronic wave functions are antisymmetric with respect to
exchanging two electrons leads to the Pauli exclusion principle, without
which electrons would all fall into the nucleus, and there would be no atoms. The exclusion principle
states that two electrons cannot be in the same state. If electrons did not
have spin the universe would be a very different place, since only one
electron could then be put into each band. Since electrons do have spin, it
is possible for two electrons to occupy the same state,
with one electron having the opposite spin of the other. In many atoms,
molecules, crystals, and liquids, the electrons are paired up, with each
member of the pair having the same probability distribution in space, but
with opposite spins. These spin states are conventionally called ``spin-up'' and ``spin-down'', 
though there is nothing ``up'' or ``down'' about them except the way they might be drawn. Note that 
one must
be careful to understand that the concept of electron pairing in states is a
rather loose way of talking. In reality electrons are indistinguishable from
each other, and one should talk about the quantum states being paired,
rather than particular electrons.

Electrons have an effective interaction that is different depending on
whether they have the same spin or not. Since electrons with opposite spin
can occupy the same space, they have a higher potential energy of interaction on average. Thus all 
else being
equal, electrons would want to have the same spin in order to lower the
system's potential energy. A system with electrons that have the same spin
direction is a ferromagnet. There is no free lunch, though. The cost of
lowering the potential energy by lining up the spins is to raise the system's
kinetic energy, since higher states must be occupied instead of the lower
energy paired states. Thus there is competition between electronic potential
energy which favors magnetism, and the electronic kinetic energy, which
favors a non-magnetic electronic structure. As pressure is increased,
electrons are pushed close together, and the relative potential energy
change between paired and unpaired electrons becomes less important; bands become wider, making the 
kinetic energy cost smaller, so that in
general materials become non-magnetic with increasing pressure. The total
energy change between the magnetic and non-magnetic system in the simple
picture presented so far is known as the exchange energy.

As temperature is raised from low temperatures, the magnetic-moment directions on each atom will
fluctuate more and more, and at some critical temperature, called the Curie
temperature, or T$_{C}$ in ferromagnets, the moments will disorder. In
general, there are still magnetic moments on the ions above T$_{C}$, they
are just disordered in direction.

Antiferromagnets have moments of opposite direction on alternating sites. It
is the different hybridization of electronic states that leads to
antiferromagnetic rather than ferromagnetic order, so that the kinetic
energy is lowered. This is sensitive to pressure, so some ferromagnets
become antiferromagnetic with increasing pressure, as in fcc iron.

In some cases the lowest free energy state has non-collinear spins. This can
arise from ``frustration,'' which is the situation where it is impossible to
tile a lattice with a perfect antiferromagnetic pattern, with every atom
having only neighbors with the opposite pointing spin. Examples are fcc and
hcp structures. Such effects can be very important to
material properties, and are responsible for the anti-Invar effect (high
thermal expansivity) in fcc Fe, for example.

This is not meant as a comprehensive review, but rather as lecture notes, and an introduction of this 
complex field to the student. Examples are mainly chosen from our own theoretical work, or from 
published experiments.

\subsection{Itinerant magnetism}

There are two endmember models for understanding antiferromagnetism. In the
band, or Slater, picture \cite{1562}, it is the different exchange interactions
between like- and unlike-spin electrons, combined with the kinetic energy of
the resulting different band states that leads to the stability of antiferromagnetic, rather
than ferromagnetic or non-magnetic states. Antiferromagnetism
is a zone-boundary instability that leads to a doubling of the
unit cell, giving folding of the electronic states. These states then
hybridize differently through the exchange potential than they would have in
the ferromagnetic case.

The Stoner model demonstrates the band picture of magnetism, and though a simple model, turns out to 
be  predictive. In the Stoner model, the magnetization energy $\Delta E$ is
\begin{equation}
\Delta E = \frac{-I M^2}{4} + \frac{M^2}{4 N(0)}
\end{equation}
where $M$ is the magnetic moment or magnetization, the Stoner Integral $I$ is an atomic property,  
$N(0)$ is the density of states at the Fermi level (or top of the valence band). The first term on 
the right hand side is the exchange energy and the second is the change in the band energy with 
magnetic moment. Minimization of $\Delta E$ gives the Stoner criterion:
\begin{equation}
I N(0) >1
\end{equation}
for a magnetic state to be stable. In the more sophisticated extended Stoner model, the average 
density of states, 
\begin{equation}
\tilde{N}(M)=M/\delta \epsilon ,
\end{equation}
is used, where $\delta \epsilon$ is the spin (exchange) splitting, and
\begin{equation}
\tilde{N}(0)=N(0)=\partial M/\partial \delta \epsilon.
\end{equation}
The instability criterion is $I \tilde{N}(M) >1$. As pressure is increased the effective density of 
states generally decreases, and magnetism will decrease and disappear with increasing pressure.

\begin{figure}[tbp]
\centering
\includegraphics[width=4in]{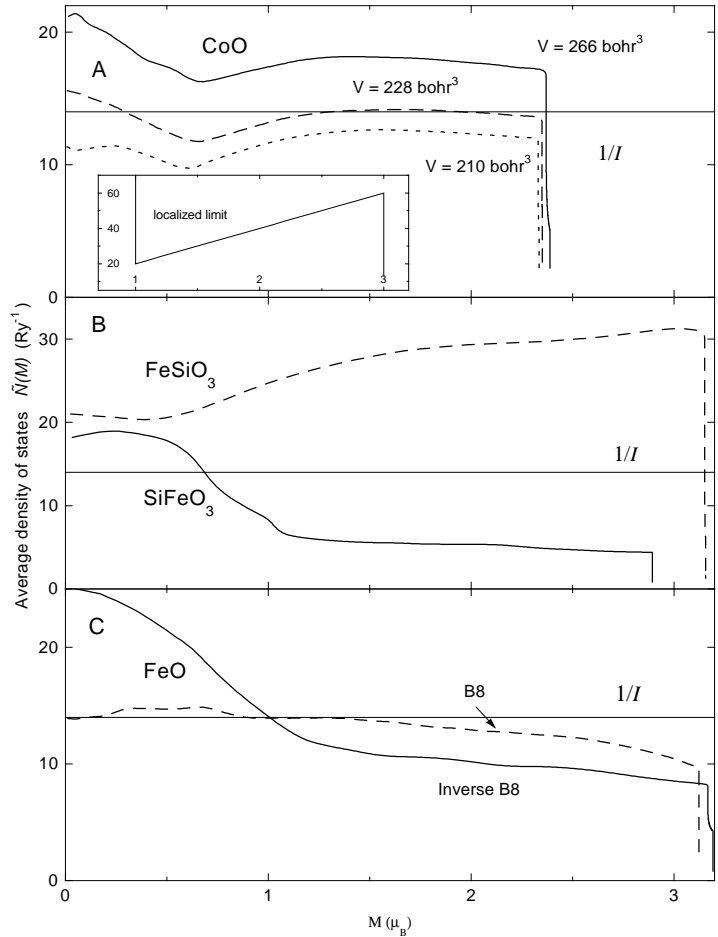}
\caption{Stoner diagram for (a) CoO in the NaCl (B1) structure with a cubic lattice, (b) FeSiO$_3$ in 
the cubic perovskite structure, and (c) FeO in the NiAs structure. The effective density of states is 
plotted 
against moment for several different volumes. The inverse Stoner parameter, 
$1/I$ is also shown, and places where the effective density of states 
crosses $1/I$ with a negative slope are stable solutions. At high volumes 
(low pressures) there is only a high-spin (large moment) solution. At 
intermediate pressures both high-spin and low-spin solutions exist. (Whether 
such coexistence is possible, i.e. whether the transition is first-order or 
second-order, depends on the shape  of the effective density of states.) At 
high pressures, there is only a low-spin solution. The Stoner model behaves 
very differently from the localized,  crystal field, picture (shown in inset) which predicts a 
discontinuous transition, never giving intermediate 
moments. From ref.~\cite{142}.}
\label{fig:stoner}
\end{figure}

Figure \ref{fig:stoner} shows Stoner diagrams for CoO, hypothetical FeSiO$_3$ in the cubic perovskite 
structure, and FeO in the NiAs structure. With increasing pressure the effective density of states 
drops, and the stable magnetic structure becomes low-spin due to this effect.

\subsection{Mott insulators}

Predicting properties of transition metal-bearing oxides
is a severe problem in modern band theory, although qualitative understanding of the behavior of 
transition metal oxides has been developed \cite{1554,1545}. Examples of problem compounds are
CoO and FeO; conventional band theory (the local density approximation, LDA,
or generalized gradient approximation, GGA \cite{500}) predicts them to be metals, but
they are actually insulators. Materials which are insulating because they
have magnetic correlations are known as Mott insulators \cite{445}. In
contrast to density functional theory (DFT) calculations within LDA or
GGA, Hartree-Fock theory gives a large band gap for transition metals oxides \cite{746}, much larger 
than
the experimentally observed gaps. It is generally understood that the the problem with
LDA and similar theories for these materials is that they underestimate the
local Coulomb repulsions between electrons. Band theory assumes that
electrons are delocalized through space, but if the electrons can be
considered as localized, the energy will depend on how many electrons are on
a given site (i.e. localized on a given atom or region of space). In order to conduct electricity the 
energy cost for an
electron to move from site to site cannot be too high. There are several
approaches that give a insulating behavior for the problematic transition
metal oxides. 

In the simplest one-band Hubbard model, the Hamiltonian is given by
\begin{equation}
E=T \sum_{ij}(c_{i\uparrow}^\dagger c_{j\uparrow} + c_{i\downarrow}^\dagger c_{j\downarrow}) + 
U\sum_i n_{i\uparrow}n_{i\downarrow}
\end{equation} 
and the band states are split into an upper and lower Hubbard band, with a splitting of $U$. The 
first term is the hopping, or hybridization, governed by $T$. The creation operator $c_i$ adds an 
electron to site $i$, and the destruction operator $c_i^\dagger$ subtracts an electron from site $i$. 
The number of up electrons on site $i$ is
\begin{equation}
n_{i\uparrow}=c_{i\uparrow}^\dagger c_{i\downarrow},
\end{equation}
and similarly for the down electrons. In this simple model there can be 0, 1 (up or down), or 2 (one 
up and one down) electrons on a site. Even this simple model has never been solved exactly in three 
dimensions. The physical picture, however, is clear. The first term is the band term. The second is a 
local repulsion. A local repulsion between electrons can open up a gap in the excitation spectrum, 
and make a insulator out of a band metal.

\subsubsection{LDA+U}

One approach that gives an insulating ground state for transition metal oxides is the LDA+U model, 
which adds a Hartree-Fock-like local Coulomb repulsion tensor $U$ and exchange interaction $J$, and 
then attempts to correct for double counting. The LDA+U model has given excellent results for a 
variety of systems \cite{25,26,1552,1553}, but the limits of the model are
still not well known, for example whether it properly predicts the high
pressure behavior of Mott insulators. Modern LDA+U is rotationally invariant, but still has some 
dependence of the choice of local orbitals in which to apply the corrections. The rotationally 
invariant LDA+U contribution to the energy is given by
\begin{eqnarray}
\Delta E &=&\frac{1}{2}\sum_{m_{1}m_{1}^{\prime }m_{2}m_{2}^{\prime }\sigma
}U_{m_{1}m_{2}m_{1}^{\prime }m_{2}^{\prime }}(n_{m_{1}m_{2}}^{\sigma }-\bar{n%
}_{m_{1}m_{2}})(n_{m_{1}^{\prime }m_{2}^{\prime }}^{-\sigma }-\bar{n}%
_{m_{1}^{\prime }m_{2}^{\prime }})+  \nonumber \\
&&+\frac{1}{2}\sum_{m_{1}m_{1}^{\prime }m_{2}m_{2}^{\prime }\sigma
}(U_{m_{1}m_{2}m_{1}^{\prime }m_{2}^{\prime }}-J_{m_{1}m_{2}m_{1}^{\prime
}m_{2}^{\prime }})(n_{m_{1}m_{2}}^{\sigma }-\bar{n}_{m_{1}m_{2}})(n_{m_{1}^{%
\prime }m_{2}^{\prime }}^{\sigma }-\bar{n}_{m_{1}^{\prime }m_{2}^{\prime }})-E_{DC} \nonumber \\
& &
\label{C1}
\end{eqnarray}
where E$_{DC}$ is the double counting correction accounting for the on-site Coulomb interaction 
already included in LDA. The tensors $U$ and $J$ are the Coulomb and exchange integrals between 
electrons in orbitals m$_i$. The $n$'s are the site occupancy matrices and $\sigma$ designates spin. 
For the double counting correction, we use the form:
\begin{equation}
E_{H}^{Model}=\frac{1}{2}\bar{U}2\bar{n}(2\bar{n}-1)-\frac{1}{2}\bar{J}[\bar{%
n}^{\uparrow }(\bar{n}^{\uparrow }-1)+\bar{n}^{\downarrow }(\bar{n}%
^{\downarrow }-1)].  \label{C6}
\end{equation}
where 
\begin{eqnarray}
\bar{U} &=&\frac{1}{(2l+1)^{2}}\sum_{mm^{\prime }}\langle mm^{\prime }|\frac{%
1}{r}|mm^{\prime }\rangle  \label{C7} \nonumber \\
\bar{J} &=&\bar{U}-\frac{1}{2l(2l+1)}\sum_{mm^{\prime }}(\langle mm^{\prime
}|\frac{1}{r}|mm^{\prime }\rangle -\langle mm^{\prime }|\frac{1}{r}%
|m^{\prime }m\rangle )  \label{C8} \nonumber \\
 & &
\end{eqnarray}
and where $\bar{n}^{\sigma }=\sum_{m}n_{mm}^{\sigma },$ and $\bar{n}=(\bar{n}%
^{\uparrow }+\bar{n}^{\downarrow })/2.$
The Coulomb and exchange tensors $U$ and $J$ are defined by 
\begin{eqnarray}
U_{m_{1}m_{2}m_{1}^{\prime }m_{2}^{\prime }} &=&\langle m_{1}m_{1}^{\prime }|%
\frac{1}{r}|m_{2}m_{2}^{\prime }\rangle  \label{C2} \nonumber \\
J_{m_{1}m_{2}m_{1}^{\prime }m_{2}^{\prime }} &=&\langle m_{1}m_{1}^{\prime }|%
\frac{1}{r}|m_{2}^{\prime }m_{2}\rangle ,
  \label{C3}
\end{eqnarray}
and are to be evaluated over localized orbitals. In practice $U$ and $J$ are input parameters, but 
can be determined from constrained occupancy computations \cite{1553,pickett}. Actually the Slater 
integrals $F_0$, $F_2$, and $F_3$ are input for $d$-states, and $U=F_0$, $J=F_2+F_4/14$, and 
$F_2/F_4=0.625$. We use the atomic values for $F_2$ and $F_3$: $F_2$=8.18 eV and $F_4$=5.15 eV, or 
$J$=0.95 eV, for CoO and $F_2$=7.67 eV and $F_4$=4.79 eV, or $J$=0.89 eV for FeO.  

LDA+U has been shown to give good predictions of the electronic structure of NiO 
\cite{1553,dudarev1,1558}. In ref.~\cite{dudarev1} the equation of state and elastic constants for 
cubic NiO were computed with LDA+U and SIC (see below), and reasonable agreement with experiment was 
found. However, there have been few tests of LDA+U total energies, and here we study in detail the 
energetics with respect to strain and compression in FeO.

\subsubsection{Self-interaction corrections}
Another successful model, which still has not been fully explored for Mott
insulators, is known as the self-interaction correction (SIC). This
is understood easiest by considering a hydrogen atom with one electron. The
self-consistent field electrostatic energy and Hartree self-consistent
potential are computed from the electronic charge density. The charge
density for a hydrogen atom is spread out, yet actually an electron is only
at one place at a time; there should be no electron-electron interaction when there is only one 
electron. The Hartree potential $V_{H}$and energy are
computed from the charge density, so that
\begin{equation}
V_{H}\left(  r\right)  =\int dr^{\prime 3}\frac{\rho
\left( r^{\prime }\right) }{\left| r-r^{\prime }\right| } ,
\end{equation} 
for example, which is an electron-electron interaction, which should not be present in a hydrogen 
atom! The
LDA exchange-correlation potential also depends on the charge density, and
would be non-zero in the hydrogen atom. In the exact density functional,
which is unknown, the exchange-correlation potential must exactly cancel the
Hartree potential, giving zero for the total electron-electron interaction.
This can be enforced, by correcting the potential so that 
\begin{equation}
V_{SIC}=V_{LDA}-\sum_{ik}\int dr^{3}V\left( \rho _{ik}\right) 
\end{equation}%
where

\begin{equation}
\rho _{ik}=\psi _{ik}^{\ast }\psi _{ik}. 
\end{equation}

The simple SIC correction above clearly depends on the localization of the
basis set. It works very well in an atom, but in a crystal the
Bloch functions are extended throughout space, and the correction goes to
zero. However, one can perform a unitary transformation on the orbitals and
find sets of localized orbitals. It has been shown that there is a
variational principle, and the most localized orbitals have the lowest SIC
energy. Thus SIC computations are quite computationally intensive, as
there is an inner iterative loop in which the orbitals are localized.
SIC does very well for excitation energies and does predict an insulating ground state for the 
transition metal oxides \cite{627,754,1556}, and gives a qualitatively correct picture for the rare 
earth metals \cite{1561}. 

In the exact density functional theory, there should be no self-interaction. One can still question 
whether the way SIC enforces freedom from self-interaction is realistic or accurate. For example, the 
weighted density approximation (WDA) is constructed to give no self-interaction per orbital, yet it 
makes only small changes in the electronic structure in most systems studied so far \cite{WDA}. In 
contrast, SIC makes large changes in electronic structure for all systems studied.

LDA+U and SIC make very different predictions for the electronic structure. Whereas LDA+U pushes 
unoccupied states up in energy, SIC pulls occupied states down in energy. This makes a difference 
when there are different types of bands present. For example, in the transition metal oxides, 4s 
states are not affected directly by $U$. The nature of the band gap is thus different in the two 
models, as will be discussed further below. In ref.~\cite{dudarev1} the EELS spectra are computed for 
O 2$p$ for LDA+U and SIC, and both compare well with each other and with experiment. However, the 
differences in models are minimized by considering only the O 2$p$ states, which are only indirectly 
affected by the correlations. 

\subsubsection{Dynamical mean field theory}
Dynamical mean-field theory (DMFT) is a method for treating correlated systems that becomes exact in 
infinite dimensions  \cite{1557}. It can be applied as an extension of LDA+U that includes a 
frequency dependent hybridization function \cite{1550}. A multiplet ion is solved self-consistently 
within the mean field Anderson impurity model. It is very time-consuming with few cases studied, and 
it is probably not currently tractable to study a system such as FeO using DMFT as a function of 
strain and pressure, as we did here. In DMFT, one still uses the parameter $U$.

\subsubsection{Hartree-Fock}
In contrast to density-functional-based methods, Hartree-Fock theory gives a large band gap (way too 
large) for transition metal oxides, and some feel that Hartree-Fock should therefore be taken as the 
zeroth order method for these materials \cite{1545,1154,746}. This is indeed the basis of the LDA+U 
method. Hartree-Fock by itself is not a reasonable way to study the high-pressure properties of Mott 
insulators, because it would grossly overestimate the pressure of a metal-insulator transition, since 
it greatly overestimates the gap. This is because the Couloumb repulsion in Haretree-Fock is 
completely unscreened. Furthermore, Hartree-Fock does not work well for any metal, always predicting 
a singularity in the density of states at the Fermi level. Thus it is not a good way to study 
metal/insulator transitions.

\section{Methods}

We have used a variety of methods to study Fe, FeO and CoO, and only a brief
outline of the methods will be given here. Computational details are given
in the referenced papers. For Fe our most accurate method is the Linearized
Augmented Plane Wave (LAPW) method. We have used the LAPW method to study
non-magnetic and collinear magnetic properties of Fe \cite{608,612,970}. LAPW is
a full-potential all-electron (i.e. no pseudopotential) method, and the
basis is very flexible, and is suited both to the interstitial region and the
atomic cores. One advantage of LAPW is that it is straightforward to
converge the results with respect to basis set size. Our computations 
are done using the Generalized Gradient Approximation (GGA) for the
exchange-correlation potential \cite{500}.

\section{Results and Discussion}

\subsection{Overview of effects of pressure on magnetism}

A straightforward effect of pressure on magnetism is through structural phase transitions. 
Compression can drive structural phase transitions, and the magnetic properties of the different 
phases will be different. For example, body-centered cubic iron is ferromagnetic, and hexagonal 
close-packed iron was believed to be non-magnetic, so the bcc to hcp transition would also be a 
ferromagnetic to non-magnetic transition. One can also have transitions from one magnetic structure 
to another. The simplest is the Curie point, where the magnetic moments become disordered. Generally, 
there are still local moments on the atoms, but they are no longer ordered. The Curie temperature in 
bcc iron is 1043 K. One could also have a transition from ferromagnetic to antiferromagnetic, or from 
an ordered structure to an incommensurate structure.

\begin{figure}[tbp]
\centering
\includegraphics[width=5in]{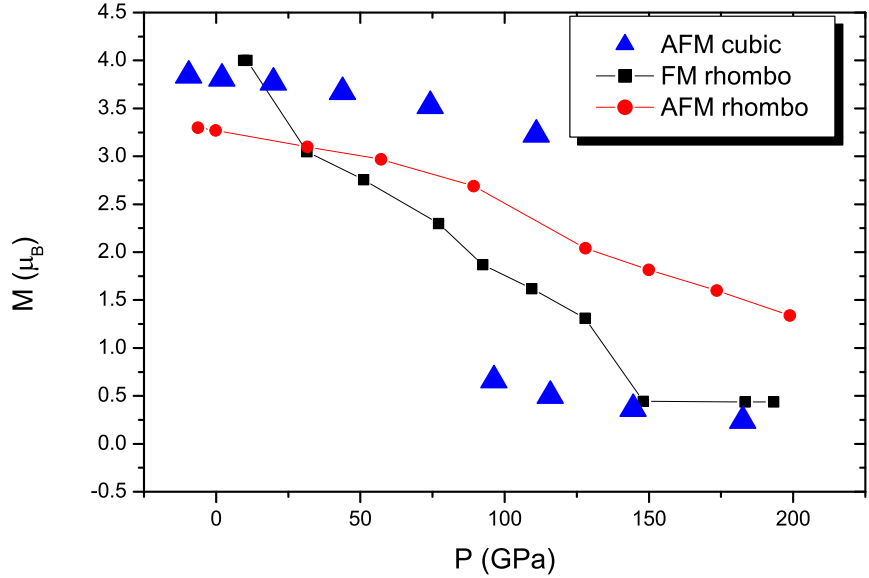}
\caption{Computed magnetic moments for FeO using the LAPW method and the GGA. For the cubic lattice 
the antiferromagnetic solution gives a discontinuous high-spin low-spin transition. However, with the 
equilibrium rhombohedral distortion at each volume, the moments decrease continuously. The behavior 
also depends on the nature of magnetic ordering, i.e. antiferromagnetic or ferromagnetic.}
\label{fig:feomoments}
\end{figure}

As pressure is increased magnetic moments tend to decrease, and eventually magnetism is squeezed out. 
This can be understood in the Stoner model as due to the general increase in bandwidths with 
pressure, decreasing the effective density of states, whereas the effective Stoner parameter is 
approximately constant. This also happens in the Hubbard picture, since $U$ is approximately constant 
or decreasing with pressure, and the hopping parameter $T$ increases in magnitude with pressure. 
Actually the parameters of the two models are closely related. Actual computations are shown for FeO 
using the LAPW method and the GGA in fig.~\ref{fig:feomoments}. Note that the magnetic behavior with 
pressure depends strongly on the crystal structure (cubic lattice {\it vs.} one with a small 
rhombohedral strain) and on the magnetic order (antiferromagnetic {\it vs.} ferromagnetic). The 
antiferromagnetic state with cubic lattice shows a strong first-order high-spin low-spin transition 
with an appreciable $\Delta V$ of 8\%. With a rhombohedral strain, or with ferromagnetic order, the 
transition is gradual. This kind of comparison is one of the useful things that can be done with 
theory. In experiments one is generally stuck with the lowest free energy phase unless there is a 
large activation energy so that metastable phases can be studied. With theory one can answer the 
question, ``What if FeO were ferromagnetic?'' High-spin low-spin transitions are predicted in all the 
transition metal oxides \cite{142, 1274}, and have been observed in sulfides and f-metals \cite{170}.

\begin{figure}[tbp]
\centering
\includegraphics[width=4in,clip]{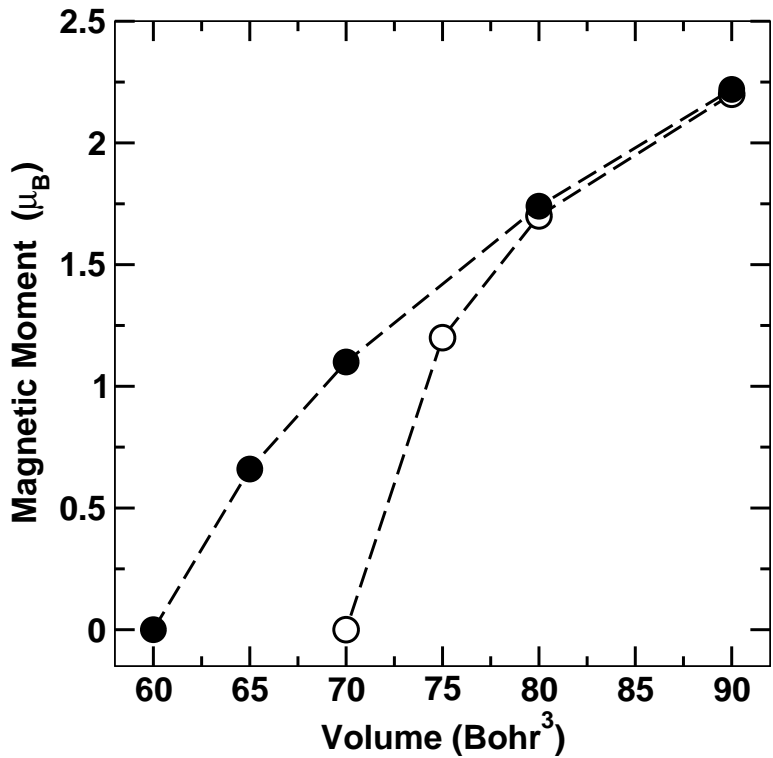}
\caption{Computed magnetic moments for afmI (open symbols) and afmII hcp Fe. These are the moments in 
the muffin tins, which have a radius of 2.0 bohr. Note that in the original presentation of these 
results (ref.~\cite{970}) the moments were plotted incorrectly smaller by a factor of two.}
\label{fig:afmmoments}
\end{figure}

Metals such as Co and hcp Fe do not show discontinuous transitions in our computations, but show a 
gradual reduction in moments as the bands widen with increasing pressure 
(fig.~\ref{fig:afmmoments})\cite{970}. The effect of pressure on magnetism is a strong function of 
the magnetic structure, as well as chemical composition.

\subsection{Magnetic behavior of Fe, FeO, and CoO with increasing pressure}

\subsubsection{Fe}

\begin{figure}[tbp]
\centering
\includegraphics[width=5in]{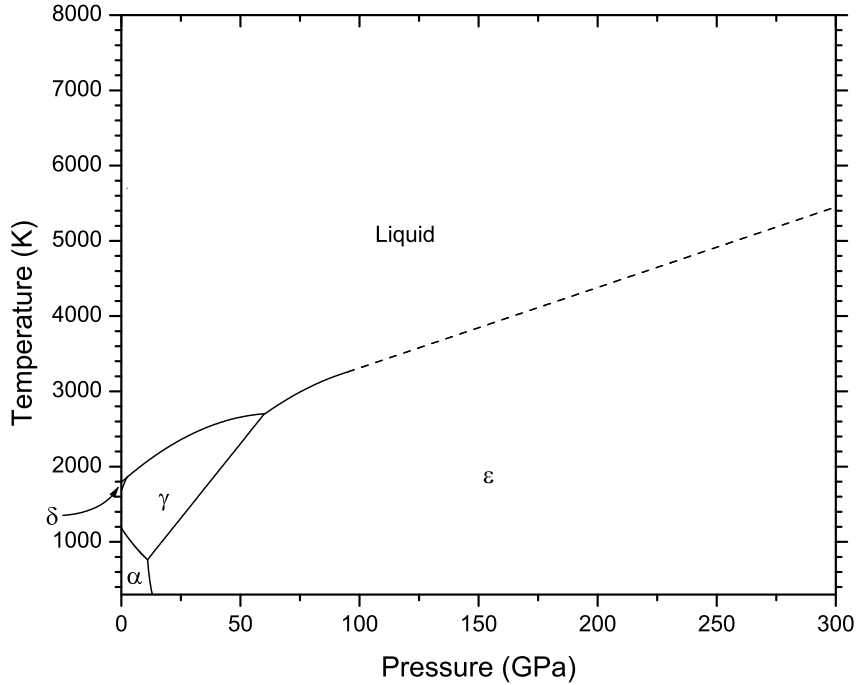}
\caption{The experimental phase diagram for Fe. See ref.~\cite{1559} for discussion.}
\label{fig:fephasediagram}
\end{figure}

There are three well-known crystalline phases for Fe: body-centered cubic (bcc, or $\alpha$-Fe), 
which is the stable form at ambient conditions, face-centered cubic (fcc, or $\gamma$-Fe) which is 
stable at high temperatures, and hexagonal close-packed (hcp, or $\epsilon$-Fe), stable at high 
pressures (fig.~\ref{fig:fephasediagram}). There is also a small bcc field ($\delta$-Fe) just before 
melting at low pressures. Bcc iron is ferromagnetic, fcc has magnetic correlations, but not an 
ordered magnetic structure, and hcp was believed to be non-magnetic (though, see discussion below). 
See ref.~\cite{1559} for a review of  experimental studies of Fe at high pressures.

One of the well-known failures of the local density approximation (LDA) is that it gives the wrong 
ground state for Fe. LDA only gives the magnetic bcc phase at negative pressures, but at zero 
pressure gives a close-packed non-magnetic ground state. One of the early successes of the GGA was 
the correct prediction of the bcc ground state and an accurate transition pressure to hcp from bcc of 
11 GPa \cite{608,1536} compared with a 10-15 GPa hysteresis loop from experiment. This showed that 
the GGA was accurate for the magnetic stabilization energy of bcc-Fe. Furthermore, bcc iron is only 
stable due to its magnetism. Non-magnetic bcc iron would not be a stable phase, except possibly in 
the $\delta$-Fe field just before melting. There had been speculations that the Earth's solid inner 
core was bcc-iron, but calculations showed that bcc-Fe was mechanically unstable at those conditions 
because it was non-magnetic at such high pressures \cite{612}.

Not only is magnetism important in bcc iron, but it is also important in fcc. We tried to find the 
phase transition from hcp to non-magnetic fcc using the particle-in-a-cell model (ref. \cite{673} and 
unpublished), but did not find a stable field for fcc below melting, which could be explained by 
magnetic stabilization of fcc. Experimentally, it is known that fcc iron is magnetic (i.e. has local 
moments).

\begin{figure}[htbp]
\centering
\includegraphics[width=5in]{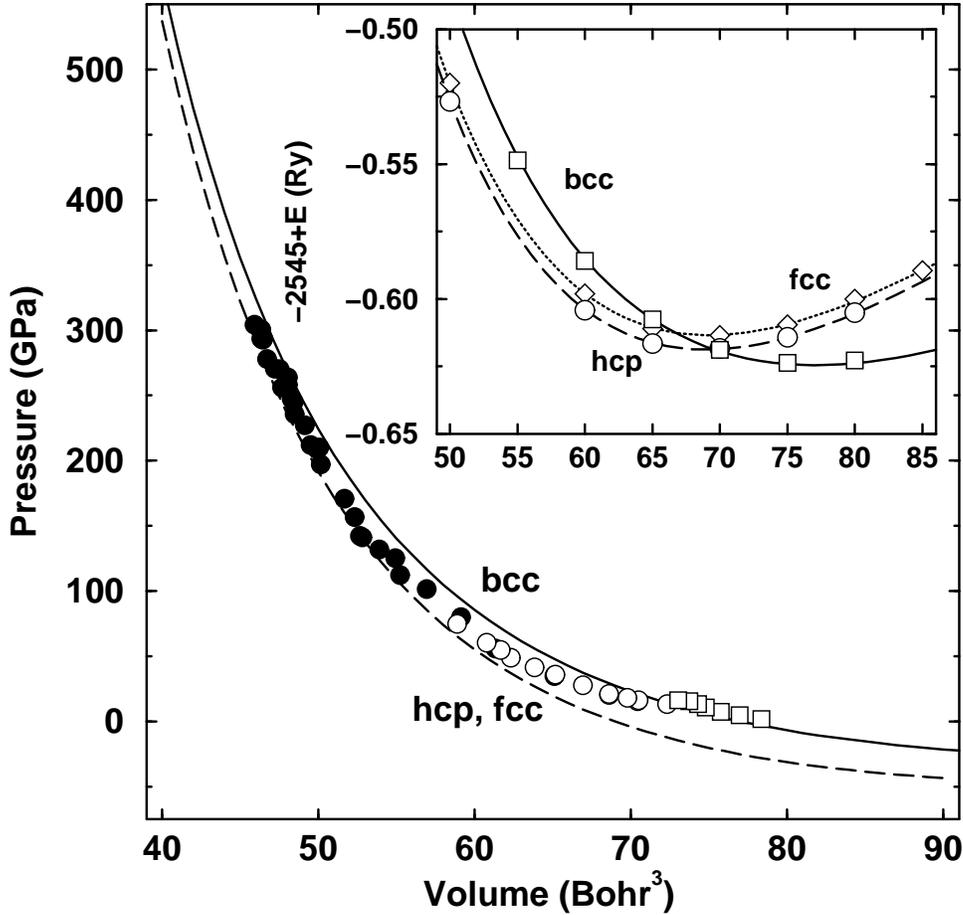}
\caption{Computed equation of state for Fe using the LAPW method and GGA. A transition pressure from 
bcc to hcp of 11 GPa is obtained from the common tangent of the energy curves shown in the inset, in 
good agreement with experiment. The computed PV equation of state is in good agreement with 
experiments \cite{1530,1531} at high pressures, and with bcc.}
\label{fig:feeos}
\end{figure}

The computed equations of state for magnetic bcc-Fe, and non-magnetic fcc and hcp, are shown in 
fig.~\ref{fig:feeos}. The inset shows the total energies for these three phases; the transition from 
bcc to hcp occurs at the common tangent. The computed P-V equation of state is in good agreement at 
high pressures, and with bcc, but discrepancies are seen at lower pressures; the non-magnetic hcp 
equation of state is too stiff at low pressures. The discrepancy at low pressures seems very large if 
one compares the zero-pressure bulk modulus K$_0$ from the equation of state. The experimental value 
is 165 GPa, compared with 292 GPa from the GGA equation of state. K$_0$ is a fictive quantity for 
hcp-Fe, since it has not proved possible to quench hcp-Fe to zero pressure, but nevertheless the 
discrepancy seems larger than expected compared with GGA results for other hcp transition metals (for 
example for Co K$_0$=190 GPa from experiment and 212 GPa from GGA, and Re is 365 GPa from experiment 
and 344 GPa from GGA \cite{970}).

\begin{figure}[htbp]
\centering
\includegraphics[width=4in]{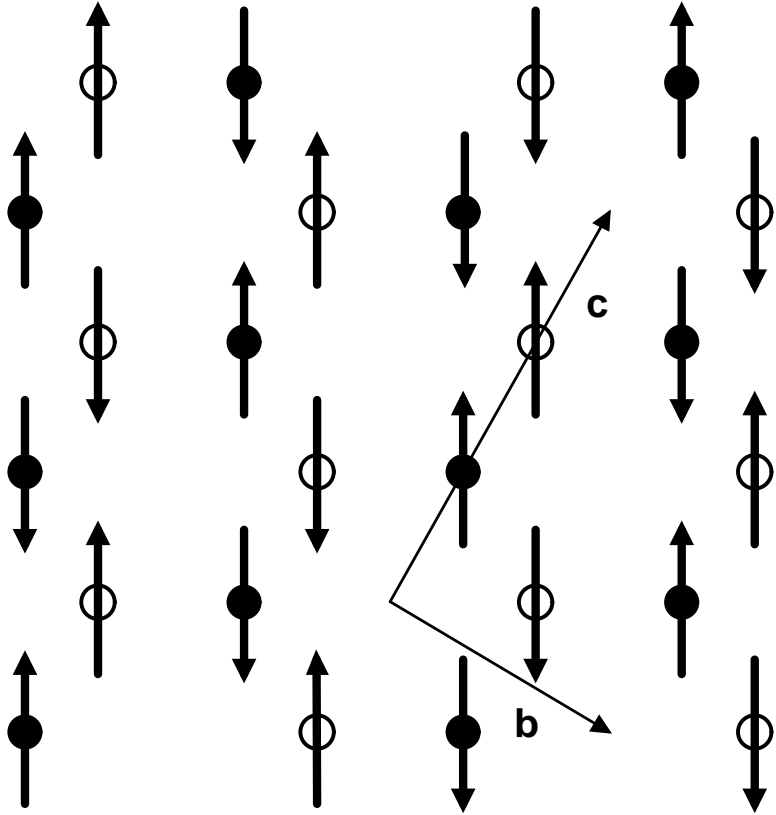}
\caption{Antiferromagnetic groundstate of hcp iron (afmII). Filled symbols show the
atomic positions at z=1/4, the open symbols at z=3/4 with the arrows
indicating the direction of spin on the atoms. The spacegroup of the afmII
structure is $Pmma$ with the atomic positions of the spin up states at
(1/4,0,1/3) and spin down states at (1/4,1/2,5/6). Also shown are the
orthorhombic unit cell vectors in the x-y-plane ($b$ and $c$): we chose
the orthorhombic $a$-axis along the hexagonal $c$-axis (out of the plane).
The orthorhombic $b$ axis then coincides with hexagonal $a$. The orthorhombic $b$ and $c$ also
define the eigenvectors for the displacements of the zone center $TO$
modes ($TO_b$ and $TO_c$) in the afmII structure.}
\label{fig:afmiistruc}
\end{figure}

Disagreements between theory and experiment often lead to advancement in theory, or in our 
understanding of physics. In this case, it appears that the discrepancy in the equation of state of 
hcp-Fe is not due to inaccuracy of the theoretical calculations or methods, but is due to the neglect 
of magnetism in hcp-Fe. Experimental data had been interpreted to show no magnetism in hcp-Fe, and 
indeed calculations showed that ferromagnetism was not stable in hcp-Fe. M\"{o}ssbauer experiments 
showed no evidence for ordered magnetism in hcp-Fe, even down to 0.03 K \cite{1533,1535}. However, 
even in the earliest papers on M\"{o}ssbauer in hcp-Fe it was recognized that magnetism was not ruled 
out by the data\cite{1543,1534}. Given the large unexplained discrepancy in the equation of state, we 
started looking for stable magnetic structures. Two stable antiferromagnetic hcp structures were 
found, and the most stable, afmII, is stable up to about 50 GPa. The computed local magnetic moments 
are shown in fig.~\ref{fig:afmmoments}. The afmI structure consists of layers of Fe with alternating 
spin along the $c$-axis. The afmII structure (fig.~\ref{fig:afmiistruc}) alternates spin-up and 
spin-down layers along the hexagonal $a$-axis \cite{970}. 

\begin{figure}[htbp]
\centering
\includegraphics[width=5in]{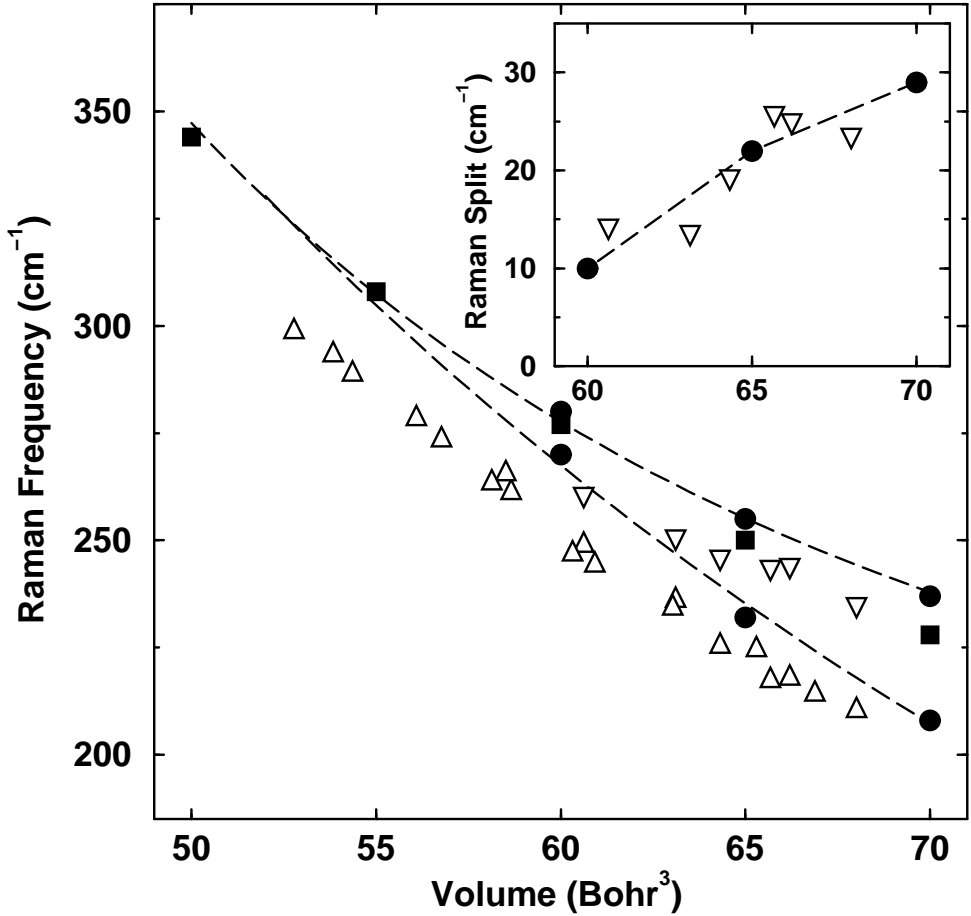}
\caption{Transverse optical frequencies as a function of atomic volume.
Non-magnetic calculations are shown in filled squares. The afmII structure
has two transverse optical modes (filled circles) with $TO_b$ being the
lower and $TO_c$ the upper branch. The dashed lines through $TO_b$ and
$TO_c$ are finite strain fits to the results to third order in
$V^{-2/3}$. Experiments \cite{1053} identify two peaks in the
Raman spectra up to 40 GPa. The stronger, low frequency peak is shown with
triangles up, the weaker, high frequency peak with triangles down. The inset
compares the calculated split in TO frequencies (circles) with the Raman
experiment (triangles).}
\label{fig:afmiiraman}
\end{figure}

The bulk modulus K$_0$ for afmII hcp-Fe is 209 GPa, a vast improvement from the non-magnetic GGA 
value of 292 GPa, but higher than the experimental value of 165 GPa. Another piece of evidence for 
local antiferromagnetic correlations comes from Raman spectroscopy. The hcp structure has a single 
Raman active mode. Raman experiments on hcp-Fe show a second broad peak \cite{1053} which cannot be 
explained within hcp symmetry. The afmII structure has magnetic symmetry lower than hcp, is 
orthorhombic, and has two transverse optic Raman modes. It is very interesting that the predicted 
splitting in Raman frequencies due to the antiferromagnetic order for afmII is in excellent agreement 
with experiment (fig.~\ref{fig:afmiiraman}). However, the second peak seen in experiments is quite 
broad, suggesting that a long-range ordered magnetic structure is not present. 

There is increasing theoretical evidence that hcp-Fe has local magnetic moments, which are not 
long-range ordered, but have strong local antiferromagnetic, non-collinear, correlations. On the 
other hand, experiments suggest that hcp-Fe is non-magnetic. M\"ossbauer data show no ordered 
moments, requiring a correlation time less than 10$^{-7}$-10$^{-9}$ seconds. X-ray absorption 
experiments \cite{1532} also show significant decrease of moment between bcc and hcp, but they do not 
prove complete loss of magnetism; the change in the absorption spectrum is due to changes in density 
of states as well as changes in the spin-related satellite. Recent experiments that show 
superconductivity in hcp-Fe \cite{1544} are fascinating. (However, the data reported in 
ref.~\cite{1544} do not prove superconductivity unambiguously; they could alternatively be 
interpreted as a magnetic phase transition. But they are strong evidence for superconductivity given 
the strong dependence of the resistivity of applied magnetic field.) It used to be thought that 
superconductivity and magnetism were incompatible, but that is now known to be not always true, and 
weak magnetism may promote exotic superconductivity \cite{1560}. Nevertheless, the fact that only 
specially treated samples were superconducting suggests that the superconducivity in Fe may be 
exotic.  The main problem with the idea that hcp Fe is locally magnetic are the M\"ossbauer 
measurements. It is hard to understand how at the lowest temperatures the spin dynamics would still 
be so rapid as to show no magnetism. On the other hand, it is rare for band theory to incorrectly 
predict a magnetic ground state. Including local correlations neglected in GGA should further promote 
magnetism. If there were structural distortions or very soft vibrational modes that invalidate the 
Born-Oppenheimer approximation there would be some grounds for considering the theoretical 
predictions less firm. But hcp is a very simple, close-packed structure, so that an incorrect 
prediction of a magnetic ground state does not seem reasonable. Furthermore, as shown above, 
including magnetism greatly improves the equation of state of Fe (table \ref{tab:fehcpeos}).  So 
there remains a problem reconciling the experimental and theoretical evidence. 

\begin{table}
\caption{Comparison of experimental and theoretical values of equilibrium 
volume ($V_0$)and bulk modulus ($K_0$) for $\epsilon$-Fe.} 
\label{tab:fehcpeos}
\begin{tabular}{cccc}
&Fe(GGA)&$V_0$ (Bohr$^3$)&$K_0$ (GPa)\\
\hline
&Expt\cite{1531,1530}&75.4&165\\
&Non-Magnetic\cite{970}&69.0&292\\
&Collinear (afmII)\cite{970}&71.2&209\\
\end{tabular}
\end{table}

\subsubsection{FeO and CoO}

Transition metal oxides like FeO and CoO present an even more challenging problem to both theory and 
experiment. Understanding materials such as FeO is one of the frontier problems in condensed matter 
physics. FeO and CoO are Mott insulators, that is they are insulating because of local magnetic 
moments. Conventional band theory makes FeO and CoO metals \cite{1547}, and no small change in 
exchange-correlation potential will make them insulators. (Large changes in the exchange correlation 
potential can make them insulators, but at the expense of accurate total energies \cite{1548}.) In 
spite of this failure, conventional LDA or GGA calculations predict energetic properties, such as 
equations of state, and the magnetic moments reasonably well \cite{268}.  LDA+U predicts a band gap 
and the canted magnetic moments and lattice strains experimentally observed in CoO are reproduced 
\cite{1549}.

We have performed LDA+U computations on CoO and FeO and get very encouraging results, but only with 
more experiments will the predictive power of LDA+U be understood. Since the results depend on the 
value of $U$, how $U$ is computed or estimated is critical. Without comparison with experiment it is 
hard to test different models for $U$, so we have used several different values. A complication of 
LDA+U is that different self-consistent results can be obtained depending on electronic symmetry and 
d-state occupations. Thus one can have electronic symmetry that is lower than the lattice or magnetic 
symmetry due to the orbital ordering. We report results on the lowest energy states we have found, 
but no systematic study of the different metastable states has yet been made.

\begin{figure}[htbp]
\centering
\includegraphics[width=5in]{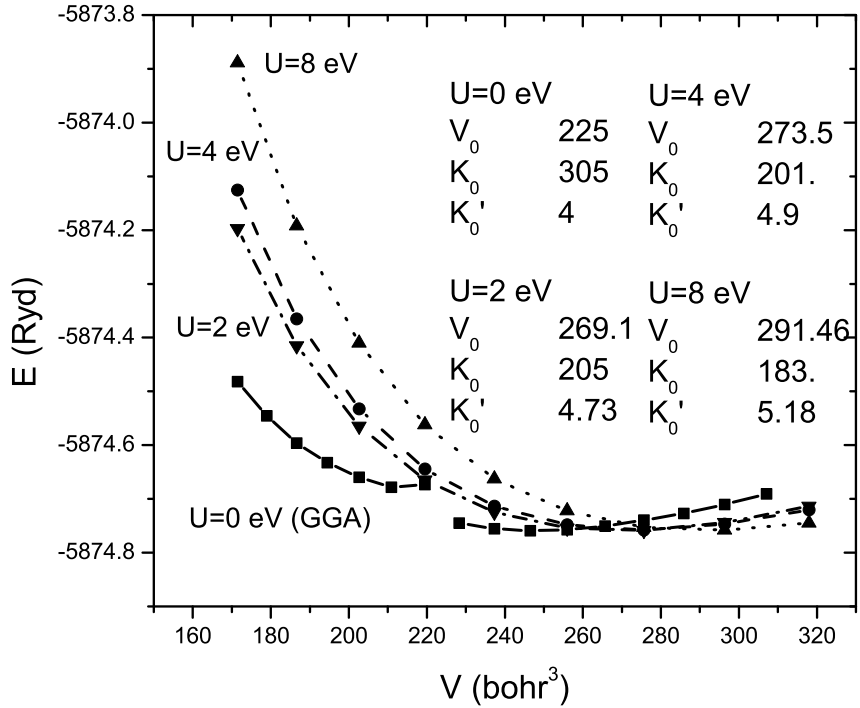}
\caption{Computed equation of state for CoO and equation of state parameters from a Vinet equation 
\cite{1010} fit to the computed energies. The experimental values for the zero pressure volume and 
bulk modulus are 261 bohr$^3$ and 181 GPa, respectively.}
\label{fig:cooev}
\end{figure}

\begin{figure}[htbp]
\centering
\includegraphics[width=5in]{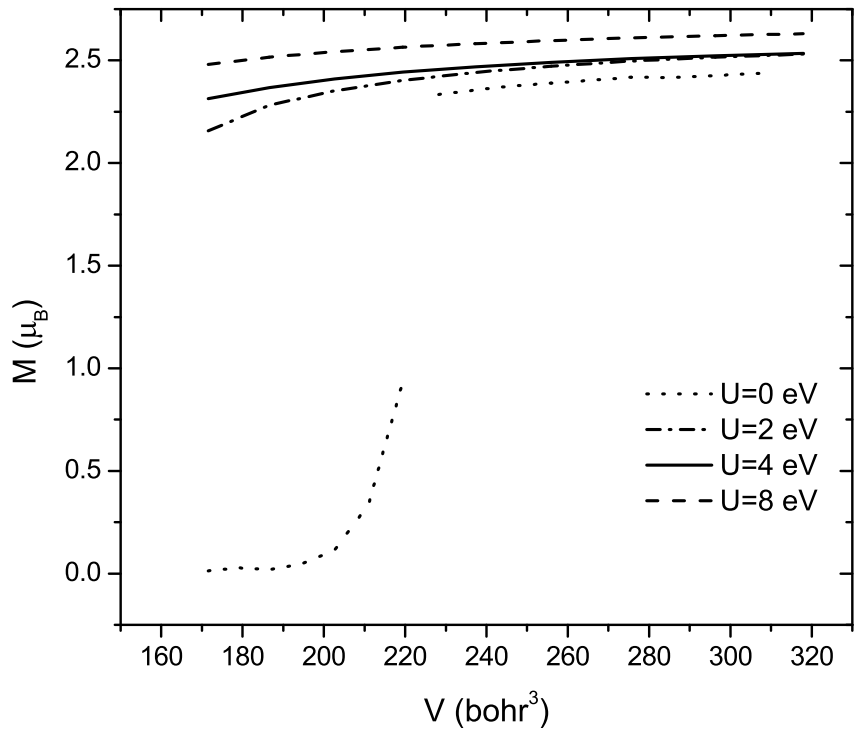}
\caption{Computed local moments for Co in cubic CoO. U=0 is the GGA result, which shows a first-order 
high-spin low-spin transition. Note that turning on U inhibits the transition.}
\label{fig:coomoments}
\end{figure}

The equations of state for CoO obtained using LDA+U (and the GGA, i.e. $U=0$) are reasonable 
(fig.~\ref{fig:cooev}), and a $U$ from 2-5 eV is consistent with the experimental equation of state. 
Figure \ref{fig:coomoments} shows the computed magnetic moments for antiferromagnetic CoO with a 
cubic lattice, computed using LMTO-ASA with GGA. The curve for $U=0$ is the GGA result. Note that our 
LDA+U results used GGA for the density functional, so our results could properly be called GGA+U. GGA 
gives a high-spin low-spin transition, as we found earlier \cite{142}. 

The band gap in CoO is predicted to initially increase with pressure, and then decrease. At zero 
pressure, the lowest conduction states are 4s states, so that the gap is an intra-atomic gap between 
3d and 4s. As pressure increases, the 4s states are driven up and the gap becomes a d-d gap, which 
then decreases with increasing pressure. LDA+U pushes unoccupied states up into the conduction band, 
so that the conduction band edge contains 4s character, which are not affected by $U$. The d-s nature 
of the gap was pointed out in ref.~\cite{mazinanisimov}; they suggested that this is consistent with 
photoemission 
data. In contrast, SIC lowers the occupied states, rather than raising the unoccupied states, and 
gives a  prediction of d-character for the lowest conduction band states \cite{754}. This suggests 
that LDA+U may be a better approximation for the nature of the gap in transition metal oxides than 
SIC. 

Of critical importance for gap closure, that is metallization pressure, is the behavior of $U$ with 
pressure. The assumption is generally made that $U$ is about constant with pressure, being a local, 
atomic-like property, but it may decrease with strong compression. In that light, the $U$=2 eV result 
of 170 GPa would be a lower bound for the metallization pressure for cubic CoO. Lattice distortion 
may increase this to higher pressure, as would a higher value of $U$. It is interesting that such 
moderate values of $U$ are sufficient to give an insulating state over a wide range of compression, 
when the GGA itself gives a metallic band structure.

\begin{figure}[htbp]
\centering
\includegraphics[width=5in]{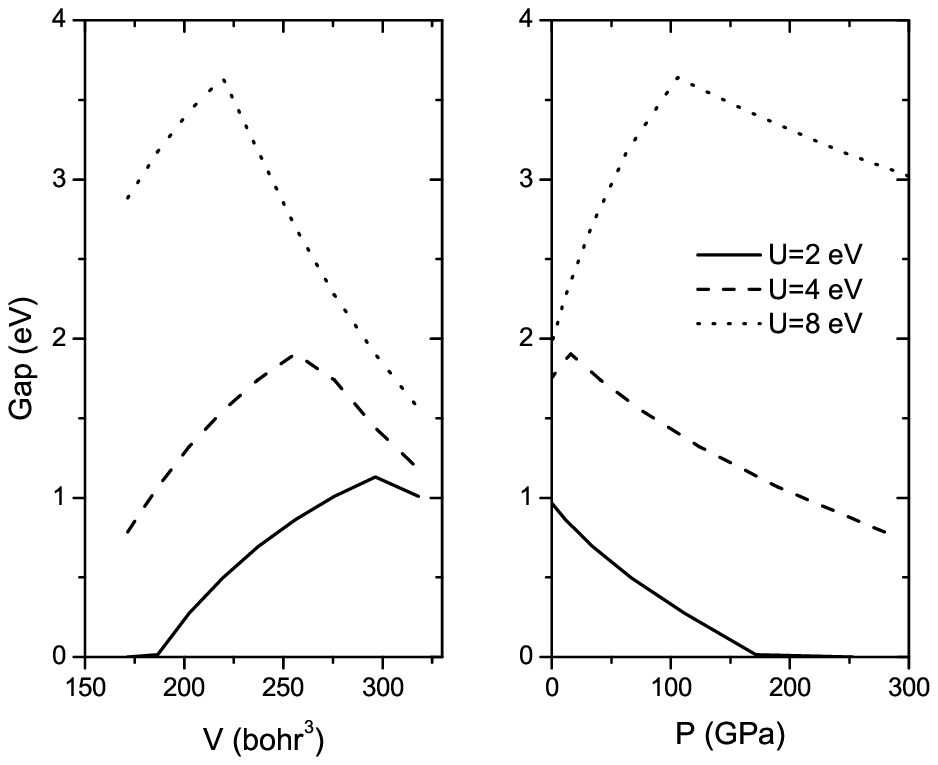}
\caption{Computed band gap for cubic CoO using LDA+U.}
\label{fig:coogap}
\end{figure}

\begin{figure}[htbp]
\centering
\includegraphics[width=5.5in]{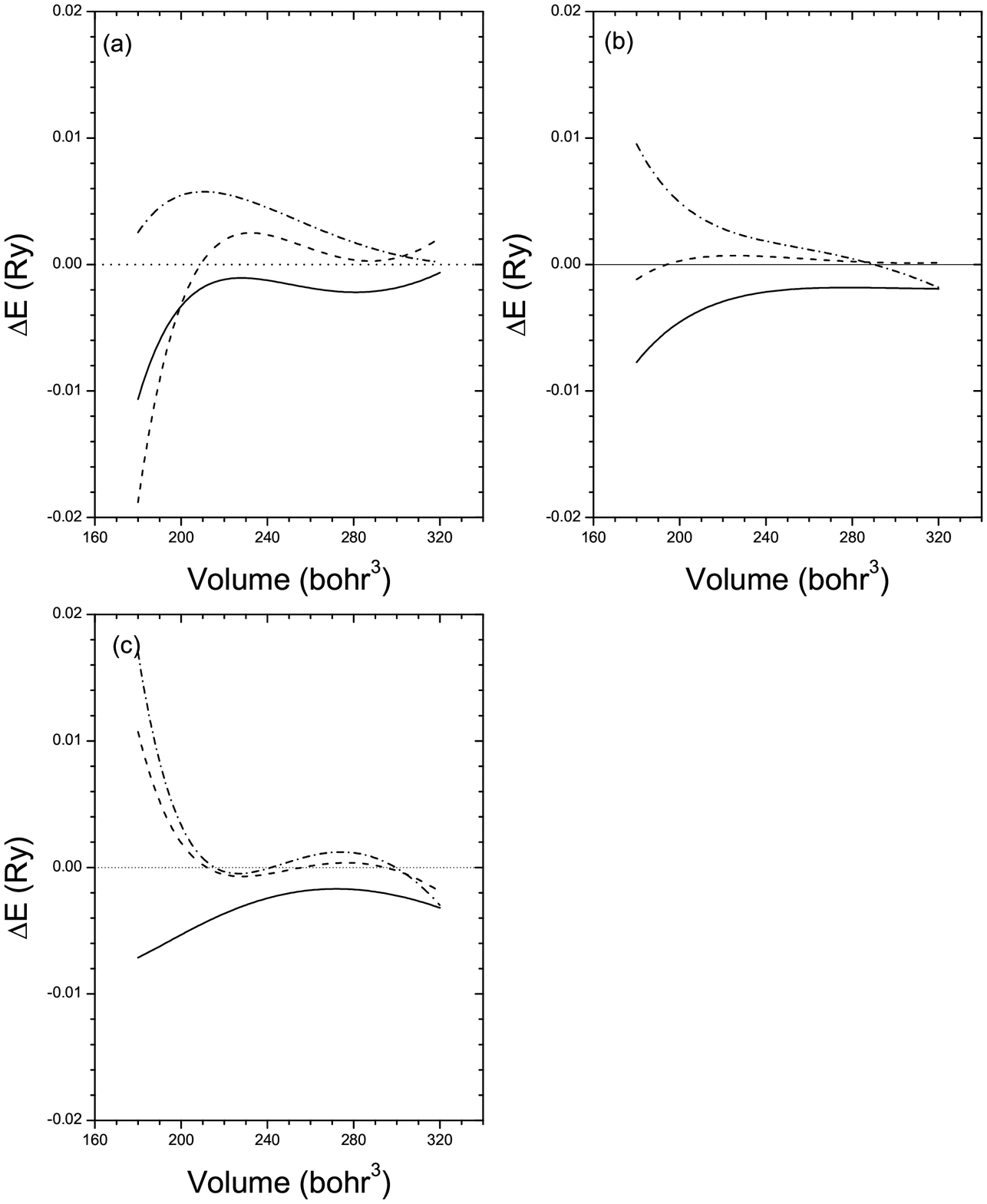}
\caption{Energy differences between various LDA + U solutions and the unstrained rhombohedral 
solution (dotted line) for (a) U = 4.6 eV, (b) U = 6.0 eV, and (c) U = 8.0 eV.  Solid line, strained 
rhombohedral solution; dashed line, strained monoclinic solution; dash-dot line, unstrained 
monoclinic solution.}
\label{fig:feodeltae}
\end{figure}

At high pressures and temperatures, FeO transforms from the rhombohedrally distorted rocksalt (B1) 
structure to a superlattice of NiAs and anti-NiAs structure \cite{191,192,409}. However, at room 
temperature, FeO can be maintained in the distorted B1 structure to over 120 GPa \cite{697}.  Here we 
discuss the behavior of FeO in the distorted rocksalt structure, since understanding the behavior of 
transition metal oxides in this simple structure is a prerequisite to understanding the behavior of 
more complex phases. We have performed a large number of self-consistent computations using the full 
potential LMTO method \cite{1550} for different lattice strains and d-orbital occupancies. The lowest 
energy state we find has rhombohedral symmetry, except for very high pressures with $U$ = 4.6 eV 
(fig.~\ref{fig:feodeltae}). This value of $U$ was obtained by computing the change in eigenvalues 
with potential shift in each d-orbital, and is probably the best estimate at zero pressure 
\cite{pickett}. As mentioned above, the effects of pressure on $U$ are still unknown. Smaller values 
of $U$ ($U \leq$ 3.5 eV) give a metallic band structure at zero pressure.

\begin{figure}[htbp]
\centering
\includegraphics[width=5.5in]{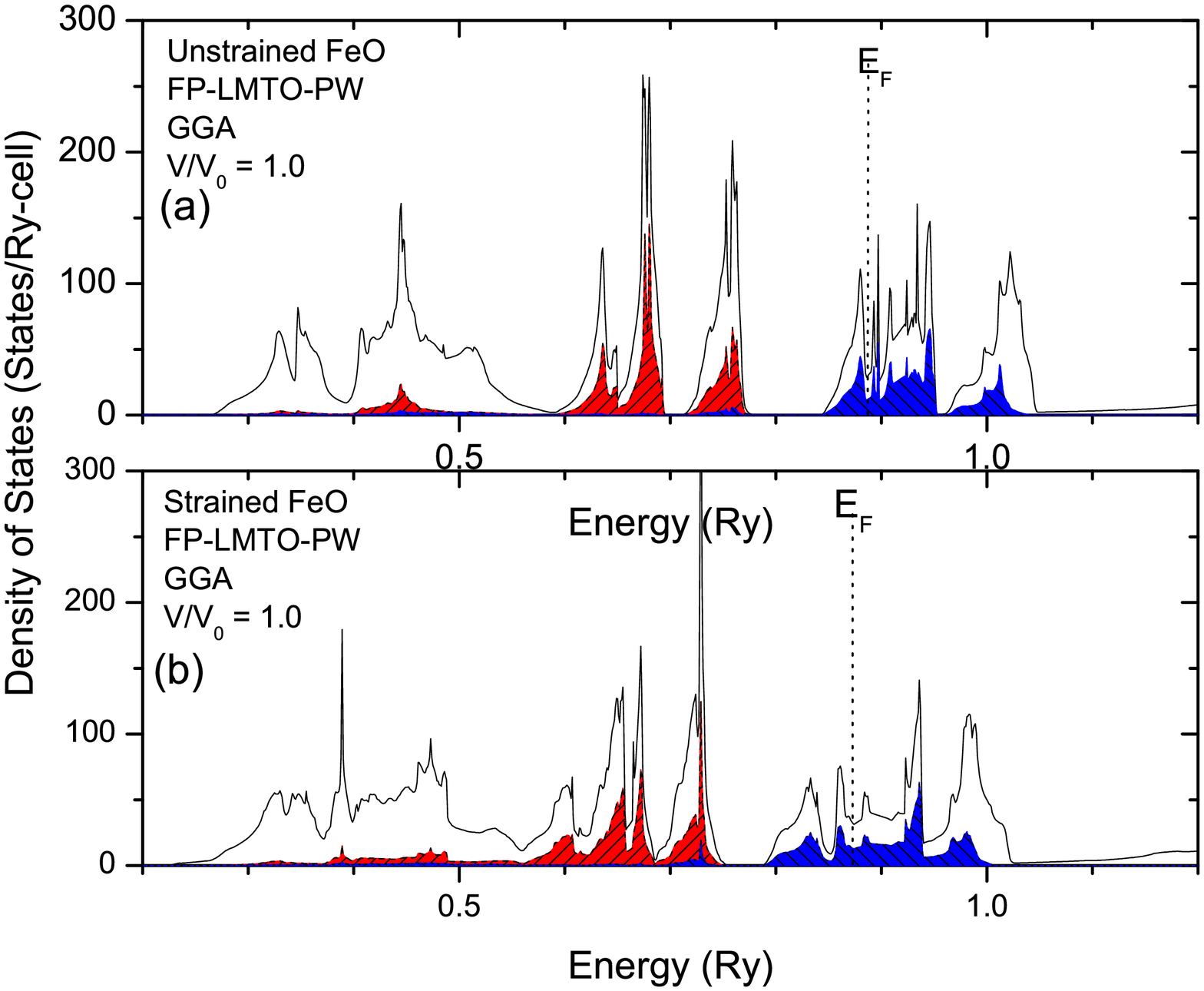}
\caption{Densities of states for FeO at the zero pressure volume for (a) cubic lattice and (b) 
optimal rhombohedral strain computed using GGA ($U=0$). Right hashing, spin-up states in one Fe 
muffin tin;  left hashing, spin-down states.}
\label{fig:dosu0}
\end{figure}

\begin{figure}[htbp]
\centering
\includegraphics[width=5.5in]{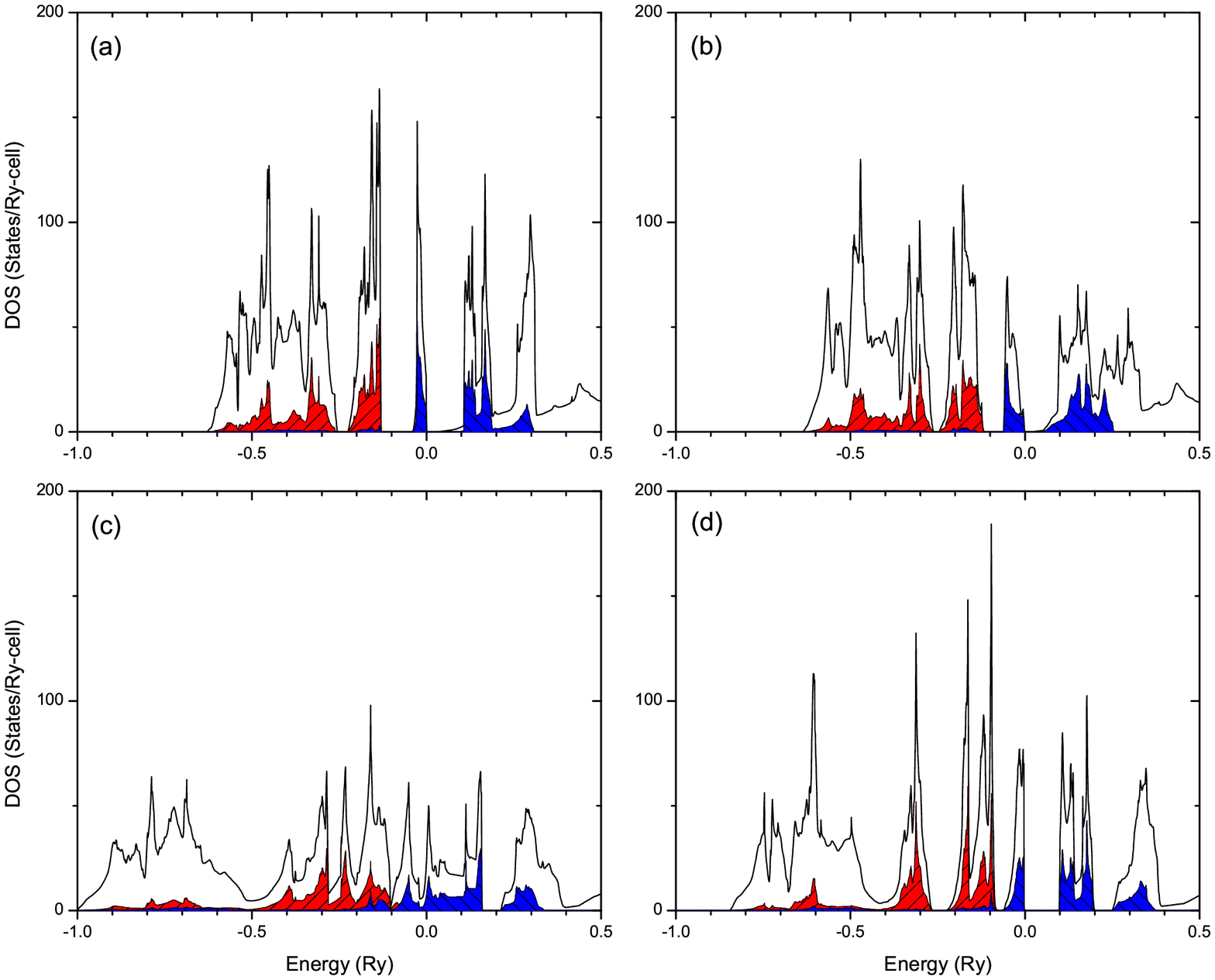}
\caption{Density of states for FeO with LDA+U, and $U$=4.6 eV.  (a) Rhombohedrally strained with 
rhombohedral electronic symmetry at 0 GPa; (b) rhombohedrally strained with electronic monoclinic 
symmetry at 0 GPa; (c) rhombohedrally strained with rhombohedral electronic symmetry at 180 GPa; (d) 
rhombohedral symmetry with a cubic lattice at 180 GPa.  Right hashing, spin-up states in one Fe 
muffin tin;  left hashing, spin-down states.}
\label{fig:feodosU46}
\end{figure}

The GGA ($U=0$) density of states is shown in fig.~\ref{fig:dosu0} at the experimental zero pressure 
volume. The electronic structure is metallic in the GGA. With a rhombohedral strain, some minority 
spin states move down, making it possible to open a gap, but still a gap does not form. LDA+U does 
open a gap (fig.~\ref{fig:feodosU46}), as was shown in the original LDA+U work (ref.~\cite{25}). 
LDA+U also does a much better job for an isolated Fe ion in MgO, indicating that it might work for 
arbitrary transition metal ion concentrations \cite{337}. 

\begin{figure}[htbp]
\centering
\includegraphics[width=5in]{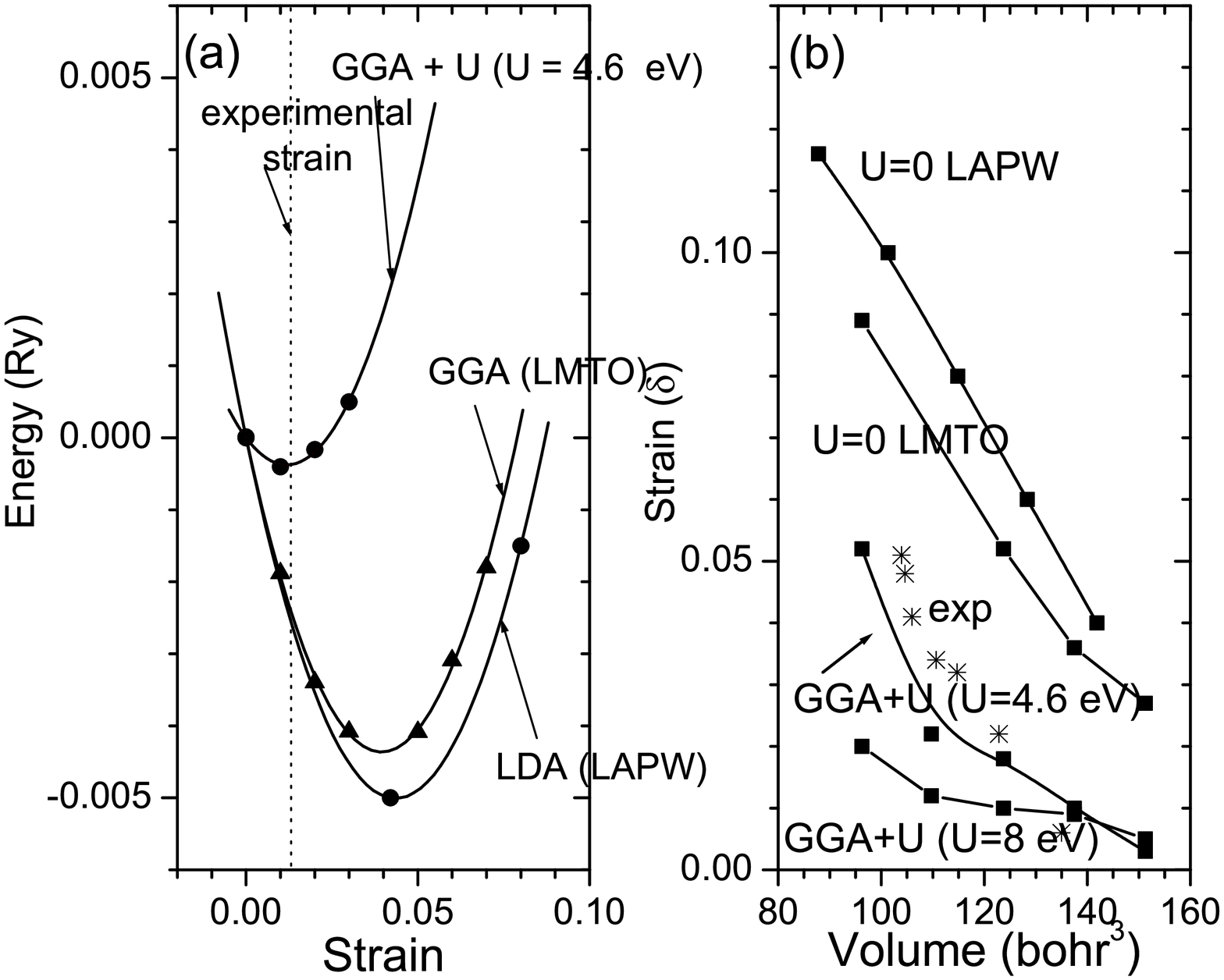}
\caption{(a) Energy {\it vs.} rhombohedral strain $\delta$ for FeO at V=137.33 bohr$^3$/formula unit, 
the experimental zero pressure volume. The LDA+U results are an improvement over GGA. Good agreement 
is found between full potential LAPW and full potential LMTO, and little difference is found between 
LDA and GGA at fixed volume. (b) Optimized strain for FeO as a function of volume.  Asterisks, 
experiment \cite{697}.}
\label{fig:strain}
\end{figure}

Previous work has not examined the LDA+U total energies. In order to understand how predictive are 
the total energies computed with LDA+U, we studied the energy as a function of rhombohedral strain 
for FeO. The lattice strain $A$ in terms of the rhombohedral strain parameter $\delta$ is given by
\begin{equation}
A=(1+3\delta)^{-\frac{1}{3}} \left[ \begin{array}{ccc} 
1+\delta & \delta & \delta \\
\delta & 1+\delta & \delta \\
\delta & \delta & 1+\delta \\
\end{array}
\right]
\end{equation}
We find that the lattice strain and its pressure dependence are predicted better by LDA+U than by GGA 
or LDA (fig.~\ref{fig:strain}). This is strong evidence that LDA+U is making the right kind of 
corrections since the energetics of strain are quite subtle.

We also found more than one stable state for FeO with different orbital occupancies. The lowest 
energy state at low pressures was the strained lattice with rhombohedral symmetry. At high pressures, 
when $U \geq$4.6 eV, we find a phase transition to a state with monoclinic electronic symmetry 
(fig.~\ref{fig:feodeltae}).

\begin{figure}[htbp]
\centering
\includegraphics[width=5.5in]{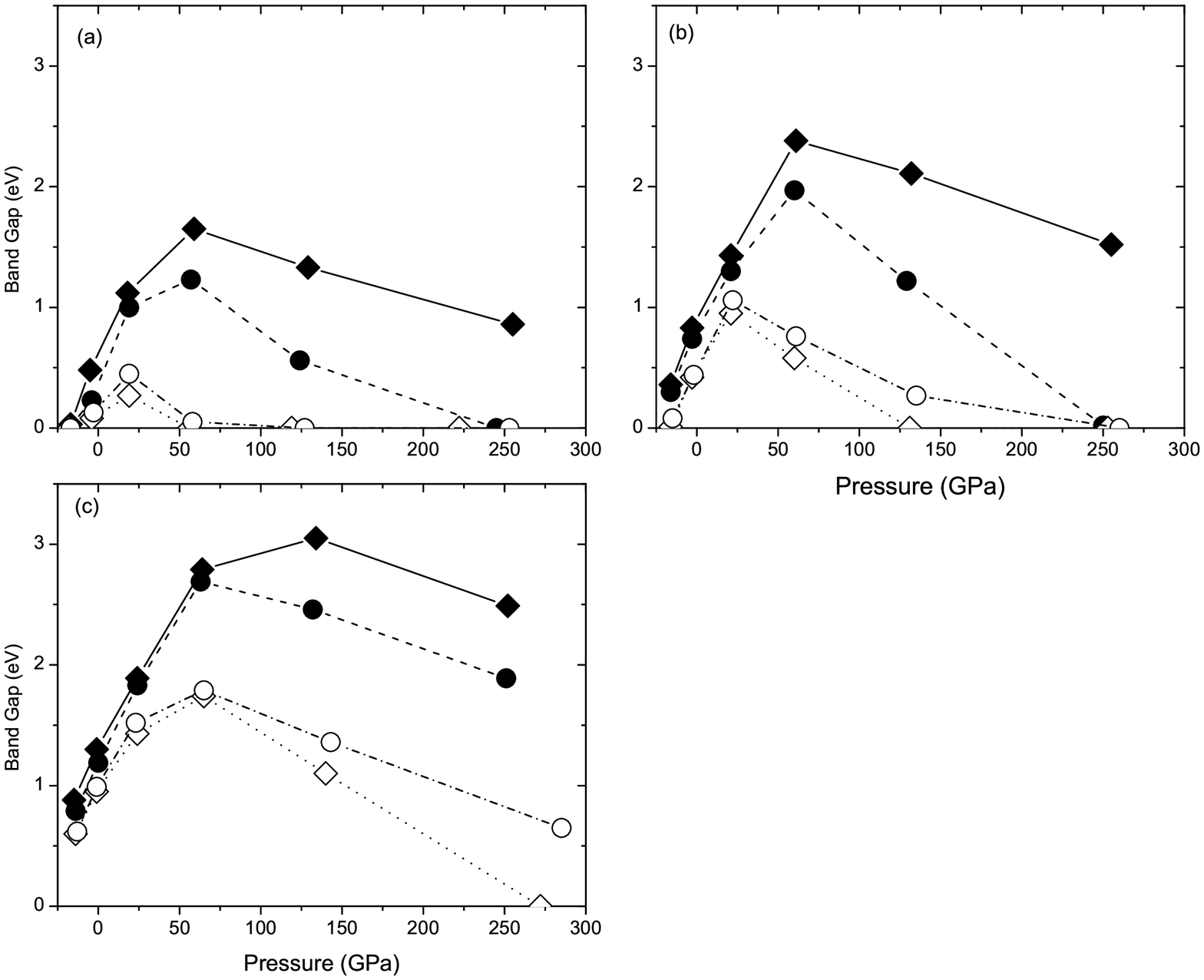}
\caption{Band gaps as a function of pressure for different LDA + U solutions for FeO:  (a) U = 4.6 
eV, (b) U = 6.0 eV, (c) U = 8.0 eV.  Solid symbols, rhombohedral solutions; open symbols, monoclinic 
solutions.  Diamonds, cubic lattice; circles, rhombohedrally-strained FeO.}
\label{fig:feogap}
\end{figure}

The band gaps for different symmetry solutions are shown in fig.~\ref{fig:feogap}. The lower energy 
rhombohedral symmetry d-state occupancies also have a larger band gap, and gap closure occurs at 
about 250 GPa for $U$=4.6 eV. The rhombohedral strain lowers the gap appreciably. It is interesting 
to note that the monoclinic structure gap closes at much lower pressures. This suggests that large 
non-hydrostatic stresses might promote gap closure at lower pressures than under hydrostatic 
conditions.

\section{Conclusions}

Magnetism is an important contribution to the high pressure properties of many materials containing 
transition metals. For pure Fe, conventional electronic structure methods within the density 
functional theory using modern exchange-correlation functionals, such as the generalized gradient 
approximation (GGA), successfully predict structural and magnetic properties, including equations of 
state, phase transition pressures, elasticity, etc. There are still some areas with discrepancies, 
such as the elastic properties of hcp Fe, but these differences are probably due both to experimental 
difficulties, and to not yet having the proper magnetic structure in the first-principles 
computations. Discrepancies in theoretical computations of the melting curve \cite{1025,1049} are 
probably due to the difficulties of sufficient accuracy in computing the solid and liquid free 
energies rather than any fundamental problem with the underlying theory. The main remaining hurdle 
for Fe is to be able to do computations with sufficient speed and accuracy, self-consistently within 
the GGA, to compute free energies of all of the phases as functions of V, T, strain, etc., in order 
to obtain a thermal equation of state, elasticity, phase transitions, magnetic, and vibrational 
properties over a wide pressure and temperature range, though much progress has been made 
\cite{1025,1049,1067,1024,1066}. Theory suggests that magnetism in Fe is key to understanding the 
high pressure behavior of Fe, for pressures at least up to 50 GPa. On the other hand, there are no 
experiments that show ambiguously the presence of local moments in hcp Fe, and many experiments imply 
the opposite. If indeed iron is non-magnetic, contrary to our best theoretical calculations, 
important changes to our best exchange-correlation functionals are required. Alternatively, perhaps 
defects and/or thermal disorder are giving rise to loss of moments, as discussed above.

The situation for transition metal oxides is not so clear. Models are available now that properly 
give insulating behavior for FeO and other transition metal oxides, such as LDA+U and SIC. It is not 
yet known how predictive these methods are since the experimental data are not yet available, though 
computations agree well with present data. Theoretical predictions of high pressure behavior within 
SIC, or within other methods such as dynamical mean field theory are also not available. Here we 
presented predictions of behavior from LDA+U that can be tested when more experimental data become 
available. In any case, it is clear that local magnetism is responsible for the insulating behavior, 
and is also probably key for accurately understanding lattice strains and elasticity.

This is a forefront area, and we expect to see many advances in both theory and experiment for the 
effects of magnetism on high pressure properties.  

\acknowledgments We thank B. Fultz, A. Goncharov, R.J. Hemley, A.I. Liechtenstein, H.K. Mao, I. 
Mazin, S. Mukherjee, S. Savrasov, and V. Struzhkin for helpful discussions. This work was supported 
by the National Science Foundation under grants EAR-9870328 (REC)), EAR-9614790 and EAR-9980553 (LS), 
and by DOE ASCI/ASAP subcontract B341492 to Caltech DOE W-7405-ENG-48 (REC). Computations 
were performed on the Cray SV1 at the Geophysical Laboratory, 
support by NSF grant EAR-9975753 and by the W.\ M.\ Keck Foundation.



\begin{thebibliography}{10}

\bibitem{825}
\BY{Mcmahan A.~K., Huscroft C., Scalettar R.~T., \atque Pollock E.~L.}
\IN{J. Comp. Aided Mat. Design}{5}{1998}{131}

\bibitem{1562}
\BY{Slater J.~C.}
\IN{Phys. Rev.}{165}{1968}{165--669}


\bibitem{1554}
\BY{Zaanen G.~J., Sawatzky A., \atque Allen J.~W.}
\IN{Phys. Rev. Lett.}{55}{1985}{418--421}

\bibitem{1545}
\BY{Brandow B.~H.}
\IN{Adv. Phys.}{26}{1977}{651--808}

\bibitem{500}
\BY{Perdew J.~P., Burke K., \atque Ernzerhof M.}
\IN{Phys. Rev. Lett.}{77}{1996}{3865--3868}

\bibitem{445}
\BY{Mott N.~F.}
\TITLE{Metal-Insulator Transitions}
(Taylor \& Francis, New York) 1990.

\bibitem{746}
\BY{Towler M.~D., Allan N.~L., Harrison N.~M., Saunders V.~R., Mackrodt W.~C., \atque  Apra E.}
\IN{Phys. Rev. B}{50}{1994}{5041--5054}

\bibitem{142}
\BY{Cohen R.~E., Mazin I.~I., \atque Isaak D.~G.}
\IN{Science}{275}{1997}{654--657}

\bibitem{25}
\BY{Anisimov V.~I., Zaanen J.,\atque Andersen O.~K.}
\IN{Phys. Rev. B}{44}{1991}{943--954}

\bibitem{26}
\BY{Anisimov V.~I., Solovyev I.~V., Korotin M.~A., Czyzyk M.~T., \atque  Sawatzky G.~A.}
\IN{Phys. Rev. B}{48}{1993}{16929--16934}

\bibitem{1552}
\BY{Liechtenstein A.~I., Anisimov V.~I., \atque Zaanen J.}
\IN{Phys. Rev. B}{52}{1995}{R5467--R5470}

\bibitem{1553}
\BY{Anisimov V.~I., Aryasetiawan F., \atque Lichtenstein A.~I.}
\IN{J. Phys.: Cond. Matt.}{9}{1997}{767--808}

\bibitem{pickett}
\BY{Pickett, W.~E., Erwin, S. \atque Ethridge, E.}
\IN{Phys. Rev. B}{58}{1998}{1201}

\bibitem{dudarev1}
\BY{Dudarev S.~L., Botton G.~A., Savrasov S.~.Y., Szotek Z., Temmerman W.~M., \atque Sutton A.~P.}
\IN{phys. stat. sol. (a)}{166}{1998}{429--443}

\bibitem{1558}
\BY{Dudarev S.~L., Peng L.~M., Savrasov S.~Y., \atque Zuo J.~M.}
\IN{Phys. Rev. B}{61}{2000}{2506--2512}

\bibitem{627}
\BY{Szotek Z.~ \atque Temmerman W.~M.}
\IN{Phys. Rev. B}{7}{1993}{4029--4032}

\bibitem{754}
\BY{Svane A. \atque Gunnarsson O.}
\IN {Phys. Rev. Lett.}{65}{1990}{1148--1151}

\bibitem{1556}
\BY{Braicovich L., Ciccacci F., Puppin E., Svane A., \atque Gunnarsson O.}
\IN{Phys. Rev. B}{46}{1992}{12165--12174}

\bibitem{1561}
\BY{Strange P., Svane A., Temmerman W.~M., Szotek Z., \atque Winter H.}
\IN{Nature}{399}{1999}{756--758}

\bibitem{WDA}
\BY{Singh, D.~J.}
\IN{Phys. Rev. B}{48}{1993}{14099-14103}

\bibitem{1557}
\BY{Georges A., Kotliar G., Krauth W., \atque Rozenberg M.~J.}
\IN{Rev. Mod. Phys.}{68}{1996}{13--125}

	
\bibitem{1550}
\BY{Savrasov S.~Y., \atque G.~Kotliar,  picture E.~Abrahams}
\IN{Nature}{410}{2001}{793--795}

\bibitem{1154}
\BY{Brandow B.~H.}
\IN{J. Alloys Comp.}{181}{1992}{377--396}

\bibitem{608}
\BY{Stixrude L., Cohen R.~E., \atque Singh D.}
\IN{Phys. Rev. B}{50}{1994}{6442--6445}

\bibitem{612}
\BY{Stixrude L.~ \atque Cohen R.~E.}
\IN{Geophys. Res. Lett.}{22}{1995}{125--128}

\bibitem{970}
\BY{Steinle-Neumann, G., Stixrude L., \atque Cohen R.~E.}
\IN{Phys. Rev. B}{60}{1999}{791--799}

\bibitem{1274}
\BY{Cohen R.~E., Fei Y., Downs R., Mazin I.~I., \atque Isaak D.~G.}
in \TITLE{High-Pressure Materials Research}, 
edited by \NAME{Wentzcovitch R., Hemley R.~J., Nellis W.~J., \atque Yu P.}
(volume 499. Materials Research Society, Pittsburgh, PA) 1998.

\bibitem{170}
\BY{Drickamer H.~G. \atque Frank C.~W.}
\TITLE{Electronic Transition and the High Pressure Chemistry and Physics of Solids}
(Chapman and Hall, London), 1973.

\bibitem{1559}
\BY{Hemley R.~J. \atque Mao H.-K.}
\IN{Intl. Geo. Rev.}{32}{2001}{1--30}

\bibitem{1536}
\BY{Singh D.~J., Pickett W.~E., \atque Krakauer H.}
\IN{Phys. Rev. B}{43}{1991}{11628--11634}

\bibitem{673}
\BY{Wasserman E., Stixrude L., \atque Cohen R.~E.}
\IN{Phys. Rev. B}{53}{1996}{8296--8309}

\bibitem{1530}
\BY{Jephcoat A.~P., Mao H.-K., \atque Bell P.~M.}
\IN{J. Geophys. Res.{1986}}{95}{21737}

\bibitem{1531}
\BY{Mao H.-K., Wu Y., Chen L.~C., Shu J.~F., \atque Jephcoat A.~P.}
\IN{J. Geophys. Res.}{95}{1990}{21737--21742}

\bibitem{1533}
\BY{Cort G., Taylor R.~D., \atque WIllis J.~O.}
\IN{J. Appl. Phys.}{53}{1982}{2064--2065}

\bibitem{1535}
\BY{Taylor R.~D., Pasternak M.~P., \atque Jeanloz R.}
\IN{J. Appl. Phys.}{69}{1991}{6126--6128}

\bibitem{1543}
\BY{Nicol M.~F. \atque Jura G.}
\IN{Science}{141}{1963}{1035--1038}

\bibitem{1534}
\BY{Taylor R.~D., Cort G., \atque WIllis J.~O.}
\IN{J. Appl. Phys.}{53}{11){1982}:8199--8201}

\bibitem{1053}
\BY{Merkel S., Goncharov A., Mao H., Gillet P., \atque Hemley R.}
\IN{Science}{288}{5471){2000}:1626--1629}

\bibitem{1532}
\BY{Rueff J.~P., Krisch M., Cai Y.~Q., Kaprolat A., Hanfland M., \etal}
\IN{Phys. Rev. B}{60}{21){1999}:14510--14512}

\bibitem{1544}
\BY{Shimizu K., Kimura T., Furomoto S., Takeda K., Kontani K., \etal}
\IN{Nature}{412}{2001}{316--318}

\bibitem{1560}

\BY{I. I. Mazin, I.I., Papaconstantopoulos, D.A. and Mehl, M.J.}
\IN{Phys. Rev. B}{65}{2002}{100511}

\bibitem{1547}
\BY{M.~R. Norman}
\IN{Phys. Rev. B}{40}{1989}{10632--10634}

\bibitem{1548}
\BY{Dufek P., Blaha P., \atque Schwarz K.}
\IN{Phys. Rev. B}{50}{1994}{7279--7283}

\bibitem{268}
\BY{Isaak D.~G., Cohen R.~E., Mehl M.~J., \atque Singh D.~J.}
\IN{Phys. Rev. B}{47}{1993}{7720--7731}

\bibitem{1549}
\BY{Solovyev I.~V., Liechtenstein A.~I., \atque Terakura K.}
\IN{Phys. Rev. Lett.}{80}{1998}{5758--5761}

\bibitem{mazinanisimov}
\BY{Mazin I.~I. \atque Anisimov V.~I.}
\IN{Phys. Rev. B}{55}{1997}{12822-12825}

\bibitem{191}
\BY{Fei Y. \atque Mao H.-K.}
\IN{Science}{266}{1994}{1668-1680}

\bibitem{192}
\BY{Y. Fei}
in \TITLE{Mineral Spectroscopy: A Tribute to Roger G. Burns},
edited by \NAME{Dyar M.~D., McCammon C.~ \atque Shaefer M.~W.}
(The Geochemical Society, Special Publication No. 5, 243-254),
1996.

\bibitem{409}
\BY{Mazin I.~I., Fei Y., Downs J.~W.~\atque Cohen R.~E.}
\IN{Amer.~Mineral.}{83}{1998}{451-457}


\bibitem{697}
\BY{Yagi T., Suzuki K., \atque Akimoto S.}
\IN{J. Geophys. Res.}{90}{1985}{8784--8788}

\bibitem{1010}
\BY{Vinet P., Rose J.~H., Ferrante J., \atque Smith J.~R.}
\IN{Phys. Cond. Matt.}{1989}{1}{1941}

\bibitem{337}
\BY{Korotin M.~A., Postnokov A.~V., Neumannn T., Borstel G., Anisimov V.~I., \atque
  Methfessel M.}
\IN{Phys. Rev. B}{49}{1994}{6548--6552}

\bibitem{1025}
\BY{Alfe D., Gillan M., \atque Price G.}
\IN{Nature}{401}{1999}{462--464}

\bibitem{1049}
\BY{Laio A., Bernard S., Chiarotti G., Scandolo S., \atque Tosatti E.}
\IN{Science}{287}{2000}{1027--1030}

\bibitem{1067}
\BY{Vocadlo L., Brodholt J., Alfe D., Price G., \atque Gillan M.}
\IN{Geophys. Res. Lett.}{26}{1999}{1231--1234}

\bibitem{1024}
\BY{Alfe D., Gillan M., \atque Price G.}
\IN{Nature}{405}{2000}{172--175}

\bibitem{1066}
\BY{Vocadlo L., Brodholt J., Alfe D., Gillan M., \atque Price G.}
\IN{Phys. Earth Planet. Inter.}{117}{2000}{123--137}

\end{thebibliography}

\end{document}